\documentclass[%
reprint,
nofootinbib,
bibnotes,
 amsmath,amssymb,
 aps,longbibliography,
]{revtex4-2}

\pdfoutput=1

\usepackage{xcolor}
\usepackage{graphicx}% Include figure files
\usepackage{dcolumn}% Align table columns on decimal point
\usepackage{bm}% bold math
\usepackage{hyperref}% add hypertext capabilities
\usepackage{cleveref}

\newcommand{\bR}{\mathbb{R}}

\begin{document}

\title{Weyl formula and thermodynamics of geometric flow}

\author{Parikshit Dutta}
 \email{parryparikshit@gmail.com}
\affiliation{%
 Asutosh College Kolkata, 92 S.P.Mukherjee Road\\
  Kolkata 700026, West Bengal, India
}%

\author{Arghya Chattopadhyay}
\email{arghya.chattopadhyay@umons.ac.be}
\affiliation{
	Service de Physique de l’Univers, Champs et Gravitation\\ Universit\'e de Mons,
	20 Place du Parc, 7000 Mons, Belgium 
}%

\begin{abstract}
We study the Weyl formula for the asymptotic number of eigenvalues of the Laplace-Beltrami operator with Dirichlet boundary condition on a Riemannian manifold in the context of geometric flows. Assuming the eigenvalues to be the energies of some associated statistical system, we show that geometric flows are directly related with the direction of increasing entropy chosen. For a closed Riemannian manifold we obtain a volume preserving flow of geometry being equivalent to the increment of Gibbs entropy function derived from the spectrum of Laplace-Beltrami operator. Resemblance with Arnowitt, Deser, and Misner (ADM) formalism of gravity is also noted by considering open Riemannian manifolds, directly equating the geometric flow parameter and the direction of increasing entropy as \emph{time direction}.

\keywords{Weyl asymptotic formula, Geometric flow}%Use showkeys class option if keyword
                              %display desired
\end{abstract}
\maketitle

%\tableofcontents

%\section{\label{sec:level1}Introduction :\protect\\ The Asymptotic distribution of Eigen values of the Laplace-Beltrami operator}

\section{Introduction:\protect\\ Weyl Asymptotic formula and entropy function}\label{sec:intro}
In the early 20th century, Hermann Weyl proved a formula for the asymptotic number of eigenvalues of the Laplace-Beltrami operator acting on space of functions satisfying Dirichlet boundary conditions on the boundary of a bounded domain $\mathbb{R}^{d}$ \cite{Weyl1911}. In particular he proved that if $N(E)$ are the number of eigenvalues of such an operator upto some value $E$, then
\begin{eqnarray}
\lim_{E\to\infty}\frac{N(E)}{E^{\frac{d}{2}}}=\frac{\omega_{d}\text{Vol M}}{(2\pi)^{d}}
\end{eqnarray}
with $\text{Vol}\,M$ being the d-dimensional volume of $M$ and $\omega_{d}$ is the volume of the unit sphere in $\bR^{d}$. He also conjectured the two term asymptotics of $N(E)$ in \cite{Weyl1913}, which was later proved by \cite{ivrii}. This can be stated as follows : Given the $d$ dimensional Laplace-Beltrami operator, the two term asymptotic formula for the number of eigenvalues $N(E)$ lying upto a given energy level $E$ is given by the following relation
\begin{eqnarray}\label{weyl1}
N(E)\sim c_{0}E^{\frac{d}{2}}\pm c_{1}E^{\frac{d-1}{2}},
\end{eqnarray}
with $M$ being a region in Euclidean space $\bR^{d}$ and `+' or `-' depends on the choice of Dirichlet or Neumann boundary conditions respectively. The constants in this case are 
\begin{eqnarray}
c_{0}=\frac{1}{(2\pi)^{d}}\omega_{d}{\text{Vol}\,}M, \quad c_{1}=-\frac{1}{4(2\pi)^{d-1}}\omega_{d-1}\text{Vol}\,\partial M,\notag\\
\end{eqnarray}
with $\text{Vol}\,\partial M$  being the (d-1)-dimensional volume of the boundary $\partial M$ and $\omega_{d},\,\omega_{d-1}$ are the volume of unit spheres in $\bR^{d}$ and $\bR^{d-1}$ respectively. One can consult \cite{Ivrii_2016} for an excellent review with further references on this topic. M. Kac \cite{Kac:1966xd} and A.Pleijel \cite{Pleijel1954ASO} investigated this matter further to ask if it is possible for an observer to decipher the shape of a membrane by \emph{listening} to its vibrations \footnote{The work of M. Kac in \cite{Kac:1966xd} is interestingly titled: \emph{can one hear the shape of a drum?}.}. It turned out that to some extent this is indeed possible as pointed out by H.P.Mckean and M.Singer \cite{Singer}. They proved the following result : let $\Delta$ be the Laplace-Beltrami operator on a $d$-dimensional Riemannian manifold $M$ without boundaries, equipped with a metric\footnote{$\Delta\equiv{1\over \sqrt{|g|}}\partial_i\left(\sqrt{|g|}g^{ij}\partial_j\right)$.} $g_{ij}$, then the partition function (heat kernel) is given by
\begin{eqnarray}\label{weyl2}
Z=\sum_{n} &&e^{-\frac{\gamma_{n}}{T}}=\frac{ T^{\frac{d}{2}}\text{Vol M}}{(4\pi)^{\frac{d}{2}}}+\frac{T^{\frac{d}{2}-1}\int\sqrt{g}dx^{d}R}{6(4\pi)^{\frac{d}{2}}}\nonumber\\
&&+\frac{T^{\frac{d}{2}-2}}{180(4\pi)^{\frac{d}{2}}}\int\sqrt{g}dx^{d}Q+\mathcal{O}(T^{{d\over 2}-3}).
\end{eqnarray}
Where $\gamma_{n}$ are the eigenvalues of $-\Delta$, and $Q=10A-B+2C$, with $A$, $B$ and $C$ being particular quadratic polynomials in the Riemann tensor \cite{Singer}, on which we will elaborate more in later sections. The idea of \emph{hearing} in this context is equivalent to inferring the area or volume of the manifold $M$ (or the \emph{shape of} $M$) from the knowledge of all the eigenvalues of $\Delta$. Equation \eqref{weyl2} therefore certainly makes the \emph{shape of $M$ audible enough}. Considering a more general case by allowing $M$ to be an open $d$-dimensional manifold with compact $(d-1)$ dimensional boundary $\partial M$, the spectra of $-\Delta$, with Dirichlet boundary conditions imposed, yield the partition function
\begin{eqnarray}\label{weyl3}
Z=&&\frac{T^{\frac{d}{2}}\text{Vol M}}{(4\pi)^{\frac{d}{2}}}-\frac{T^{\frac{d-1}{2}}\text{ Vol} \partial\text{M}}{4(4\pi)^{\frac{d-1}{2}}}+\frac{T^{\frac{d}{2}-1}\int\sqrt{g}dx^{d}R}{6(4\pi)^{\frac{d}{2}}}\notag\\
&&-\frac{T^{\frac{d}{2}-1}\int\sqrt{g'}dx^{d-1}H}{6(4\pi)^{\frac{d}{2}}}+O(T^{\frac{d-3}{2}}),
\end{eqnarray}
with $g'_{ij}$ being the metric on the boundary of $M$ i.e. $\partial M$ and $H$ being the mean curvature on $\partial M$. Therefore $M$ is again \emph{audible} along with the shape of the boundary $\partial M$.

 It is interesting to note a bridge between statistical physics and geometric quantities through these formulas by considering the eigenvalues of $\Delta$ as energy eigenstates of some statistical system. Although not pertaining to Laplace-Beltrami operator, the idea of geometric flows originating from the minimisation of energy functionals can be traced back to the main ideas behind Ricci flow in \cite{Eells1964HarmonicMO,Hamilton}. The author  of \cite{Hamilton} started with a metric $g_{ij}$ with strictly positive Ricci curvature $R_{ij}$ and proposed \emph{improvement} of the metric by the means of a heat equation known as the Hamilton's Ricci flow equation. Interestingly G. Perelman in \cite{perelman2002entropy}, was the one to show that the Ricci flow equation can be formulated as an energy minimisation problem through the Perelman $\mathcal{F}$ functional, which can also be thought of as a string model in dilaton gravity from the physicists perspective. This was in fact the key step in proving the Poincar\'e conjecture. Ricci flow has also been shown to be the geometric counterpart of the renormalisation group flow for string sigma models in \cite{Woolgar:2007vz,Carfora:2010iz}.

 Ricci flow has been studied in various contexts of physics such as different saddle points or blackhole solutions of 4 dimensional gravity\cite{Headrick:2006ti}, study of the formation of singularity in 3 dimensions\cite{Garfinkle:2003an} or 2 dimensional flows asymptoting to dilaton black holes \cite{Hori:2001ax}. The other forms of geometric flows barring the Ricci flow has found less space in the physics literature.
 In this paper, our goal is to establish an equivalency between entropic and geometric extremisation problems. This is done by constructing suitable entropy functions, and understanding geometric flows as positive entropy flow of the system. We found relations between the area preserving curve shortening flow of \cite{Gage} or the volume preserving mean curvature flows shown in \cite{huisken1984flow,Huisken+1987+35+48}, with the law of increment of entropy or second law of thermodynamic. In the following we will first discuss equations \eqref{weyl2} and \eqref{weyl3} from the perspective of statistical ensembles in \cref{sec:ensembles}. \Cref{sec:geom} will elaborate how this two ideas of geometric flows and entropy functions are actually related. Finally we will end with a discussion of results in \cref{sec:discussion}.

\section{ The Ensembles }\label{sec:ensembles}
To establish the relation between geometric flows and the direction of increasing entropy, first we need to define suitable ensembles. In the following we look at two such possibilities.

\subsection{Microcanonical Case}
Consider an ensemble composed of a system, whose energy values are given by the eigenvalues of the Laplace-Beltrami operator. This can be thought of as a particle moving inside this bounded domain $M$. Then the number of energy states lying between $[E,E+dE]$ is given by 
\begin{eqnarray}
\nu(E)dE=\frac{dN}{dE}dE,
\end{eqnarray}
with $N(E)$ being equivalent to (\ref{weyl1}), for large values of $E$. Thus one can define a microcanonical ensemble, consisting of all the systems which lie in this energy range, with entropy 
\begin{eqnarray}\label{eq:SB}
S_{B}=\ln(\nu(E) dE).
\end{eqnarray}
This directly follows from the definition of microcanonical ensemble for continuous energy spectra. One should not the crucial dependence of \ref{eq:SB} on the energy width $dE$. For a finite but arbitrarily small $dE$, this is the entropy associated with the microcanonical ensemble. For fixed $dE$, we can analogously define the surface entropy as
\begin{eqnarray}\label{entropymicro}
S_{surf}=\ln(\nu(E)).
\end{eqnarray}
\subsection{Canonical case}
On the other hand, consider (\eqref{weyl2}) and (\eqref{weyl3}), which is a canonical partition function. One can define the Gibbs entropy via these partition functions as 
\begin{eqnarray}\label{entropycanonical}
S_{Gibbs}=\frac{\partial [T\ln Z]}{\partial T},
\end{eqnarray}
and the free energy as
\begin{eqnarray}
U=T^{2}\frac{\partial \ln Z}{\partial T},
\end{eqnarray}
where the volume of the system is supposed to be kept fixed.

In both of the above cases, we obtain an entropy-function depending on the geometric properties of $M$, which leads to an equivalency between entropic and geometric extremization problems.

\section{Geometric flow }\label{sec:geom}
Since the ensembles themselves depend on the geometric quantities, one can imagine a set of geometric configurations in $d$ dimensional space, where each configuration can be understood as a continuous deformation from the previous one with respect to some parameter $t$. These deformations can then be continuously labelled by increasing values of the entropy function $S$. Thus the geometric flows are defined by the condition $\displaystyle{dS/dt>0}$, with respect to some global parameter $t$, which one might be inclined to call \emph{time}. Further justification for which is elaborated in \cref{subsec:OpenMC}.

\subsection{Area preserving curve shortening flow}
Consider the microcanonical case, defined in previous section, where we take the domain to be bounded by a simply connected closed curve in the two dimensional plane. Let the coordinates of the curve $C$ be $X=(x(u),y(u))$, parametrized by $u\in[0,1]$. The length of the curve $L$ (or the perimeter) and the area $A$ enclosed by it can then be written as 
\begin{eqnarray}
L&&=\int_{0}^{1}\sqrt{(dx(u))^{2}+(dy(u))^{2}}\notag\\
&&=\int_{0}^{1}\sqrt{\sum_{i=1}^{2}\bigg(\frac{dx_{i}(u)}{du}\bigg)^{2}}du=\int_{0}^{1}|X_{u}|du,
\end{eqnarray}
and 
\begin{eqnarray}
&&A=\frac{1}{2}\oint_{C}\left[xdy-ydx\right]\notag\\
&&=\frac{1}{2}\int_{0}^{1}\left[x(u)\frac{dy(u)}{du}-y(u)\frac{dx(u)}{du}\right]du\notag\\
&&=\frac{1}{2}\int_{0}^{1}\epsilon_{ij}x_{i}(u)\frac{d x_{j}(u)}{du}du.
\end{eqnarray}
The entropy in (\ref{entropymicro}), can then be written using $A$ and $L$ as
\begin{eqnarray}
S_{surf}\approx\ln&&\bigg(\frac{d\,\omega_{d}}{2(2\pi)^{d}}E^{\frac{d-2}{2}}\text{Vol} M\bigg)\notag\\
&&-\frac{2\pi(d-1)\omega_{d-1}}{4d\,\omega_{d}}E^{-\frac{1}{2}}\frac{\text{Vol} \partial M}{\text{Vol} M},
\end{eqnarray}
where we have used \eqref{weyl1} with the Dirichlet boundary condition. Applied to $d=2$ case, $\text{Vol} M\equiv A$, $\text{Vol }\partial M\equiv L$, $\omega_{2}=\pi$ and $\omega_{1}=2\pi$. Clearly this Entropy increases as the boundary volume $\text{Vol } \partial M$ or $L$ decreases, and as $\text{Vol} M$ or $A$ increases. It is suitable to keep one of them fixed (condition for microcanonical ensemble), for the analysis. Then for constant $A$ we have
\begin{eqnarray}
\frac{dS_{surf}}{dt}=-\frac{\pi}{2\,A}E^{-\frac{1}{2}}\frac{dL}{dt};\qquad 
\frac{dA}{dt}=0.
\end{eqnarray}
The change in the perimeter $L$ and the area $A$ can be evaluated directly as
\begin{eqnarray}
&&\frac{dL}{dt}=\int_{0}^{1}\frac{X_{u}.\, \partial_{t}X_{u}}{|X_{u}|}du=-\int_{0}^{L}\partial^{2}_{s}X.\partial_{t}X\,ds,\\
&&\frac{dA}{dt}=\int_{0}^{1}\epsilon_{ij}\partial_{t}x_{i}(u)\frac{dx_{j}(u)}{du}du,
\end{eqnarray}
where $``."$ represents the normal two dimensional vector dot product and $ds$ the infinitesimal arc-length. The curvature dependence of the flow is clear from appearance of the second derivative $\partial^{2}_{s}X$. Solution to this problem is well known, and is called the area preserving curve shortening flow \cite{Gage}. 
\begin{eqnarray}
\frac{\partial X}{\partial t}=-\bigg(\kappa-\frac{2\pi}{L}\bigg)\mathcal{N},
\end{eqnarray}
where $\mathcal{N}$ is the unit outward normal, and $\kappa$ is the Euclidean curvature. A variation of this flow was also proposed in \cite{Gage}. The higher dimensional generalizations of this are known as the \emph{Mean curvature flow} and the \emph{volume preserving Mean curvature flow} (\cite{huisken1984flow,Huisken+1987+35+48}). We will elaborate more on these in section \ref{subsec:OpenMC}. The famous Ricci flow problem is also closely related to this. Let $\text{Vol M}=\frac{1}{d}\int(\vec{r}.\hat{d})\sqrt{g}dx^{d-1}=\int e^{-f}\sqrt{g}dx^{d-1}$ be the volume enclosed by a $(d-1)$ dimensional hypersurface, with $\text{Vol } \partial\text{M}=\int\sqrt{g} d^{d-1}x$ being the boundary volume. To keep the volume invariant, we must have the condition
\begin{eqnarray}
\delta\text{Vol M}=\frac{1}{d}\int e^{-f}\sqrt{g}dx^{d-1}(-\delta f+\frac{1}{2}g^{ij}\delta g_{ij}).
\end{eqnarray}
While variation of the boundary volume yields
\begin{eqnarray}
\delta\text{Vol } \partial\text{M}=\int \sqrt{g}dx^{d-1}\frac{1}{2}g^{ij}\delta g_{ij}.
\end{eqnarray}
There are two intuitive ways to get a geometrization via increasing entropic flow. 
\begin{itemize}
\item If the variation of the metric is chosen to be proportional to the metric as $\delta g_{ij}=dt\,\partial_{t}g_{ij}=dt\,p\,g_{ij}$, such that 
\begin{eqnarray}
\delta\text{Vol } \partial\text{M}=dt \frac{d}{2}\int \sqrt{g}dx^{d-1}p<0.
\end{eqnarray}
Choosing $p=q(\langle q\rangle-q)$ for some scalar function on the boundary $\partial \text{M}$ then yields
\begin{eqnarray}
\delta\text{Vol } \partial\text{M}=-dt \text{Vol}\partial \text{M}\frac{d}{2}(\langle q^{2}\rangle-\langle q\rangle^{2})<0.
\end{eqnarray}
The choice of equating $q$ with the scalar curvature of the surface $R$, closely relates this to Yamabe Flow as proposed by R. Hamilton. 

\item One may also choose $\delta g_{ij}=dt\,\partial_{t}g_{ij}=dt\,(\langle m\rangle-m)m_{ij}$, where $m_{ij}$ is some second rank tensor and $m=g^{ij}m_{ij}$. This yields
\begin{eqnarray}
\delta\text{Vol } \partial\text{M}=-\text{Vol}\partial \text{M}\frac{dt}{2}(\langle m^{2}\rangle-\langle m\rangle^{2})<0.
\end{eqnarray}
Choosing $m_{ij}=h_{ij}$, the second fundamental form of $\partial \text{M}$, relates this to the volume preserving Mean curvature flow. We will come back to this later. 
\end{itemize}

The volume constraint on the other hand is imposed by 
\begin{eqnarray}
\frac{\partial f}{\partial t}=\frac{1}{2}g^{ij}\frac{\partial g_{ij}}{\partial t}.
\end{eqnarray}
In fact, there are several other choices that one can make. One particular case relates the entropic flow to Perelman's functional. If the variation of the metric is chosen to satisfy
\begin{eqnarray}
\delta g_{ij}=dt\frac{\partial g_{ij}}{\partial t}=-dt e^{-f}2(R_{ij}+\nabla_{i}\nabla_{j}f)
\end{eqnarray}
Then the entropy changes as
\begin{eqnarray}
&&\frac{\partial S_{surf}}{\partial t}\approx\frac{2\pi(d-1)\omega_{d-1}}{4d\,\omega_{d}\text{Vol M}}E^{-\frac{1}{2}}\int e^{-f}(R+|\nabla f|^{2})\sqrt{g} d^{d-1}x,\notag\\
\end{eqnarray}
which is exactly the Perelman $\mathcal{F}$ functional \cite{perelman2002entropy}. One can show that the derivative of this is positive definite following the same arguments as Perelman by showing
\begin{eqnarray}
&&\frac{\partial^{2} S_{surf}}{\partial t^{2}}=\frac{\partial\mathcal{F}}{\partial t}\approx\notag\\
&&\approx \frac{2\pi(d-1)\omega_{d-1}}{4d\,\omega_{d}\text{Vol M}}E^{-\frac{1}{2}}\int e^{-2f}|R_{ij}+\nabla_{i}\nabla_{j}f|^{2}\sqrt{g} d^{d-1}x,\notag\\
\end{eqnarray}
implying the monotonic increment of  $S_{surf}(t)$.

\subsection{Curvature related flows}
In the previous microcanonical case, we check that geometric flow equations naturally satisfy the entropy condition, $\frac{d S}{dt}>0$, but they are not derived directly using the entropy as a Dirichlet energy functional. The reason of which being the absence of any curvature dependence in the two term asymptotic formula for $N(E)$. On the other hand in the canonical case, curvature dependent terms appear naturally. It is thus more interesting to study geometric flows arising from the positivity of the entropy function in such a case. 

\subsubsection{Closed manifold}
Consider the canonical ensemble, and the Gibbs entropy function (\ref{entropycanonical}) evaluated from the partition function (\ref{weyl2})
\begin{eqnarray}
&&S_{Gibbs}=\notag\\
&&\frac{\partial}{\partial T}\bigg[T\ln\bigg(\frac{T^{\frac{d}{2}}\text{Vol} M}{(4\pi)^{\frac{d}{2}}}\bigg)+\frac{1}{6}\frac{\int \sqrt{g}dx^{d} R}{\text{Vol} M}\\
&+&\frac{1}{180 T}\frac{\int\sqrt{g}dx^{d}Q}{\text{Vol} M}-\frac{1}{72T}\bigg(\frac{\int \sqrt{g}dx^{d} R}{\text{Vol} M}\bigg)^{2}+O\left({1\over T^2}\right)\bigg].\notag
\end{eqnarray}
Since the generic form of $Q$ in dimensions greater than $2$ is more complicated, we stick to dimension 2 for our discussion. In that case integral of the scalar curvature is just $4\pi$ times the Euler character, being a constant for the manifold, while $Q=3R^{2}$ \cite{Singer}. The change of the entropy with time then yields
\begin{eqnarray}
&&\frac{\partial S_{Gibbs}}{\partial t}\approx\notag\\
&&-\frac{1}{60T^{2}\text{Vol} M}\int\sqrt{g}dx^{2}\bigg(\frac{1}{2}R^{2} g^{ij}\partial_{t}g_{ij}+2R\partial_{t}R\bigg).\notag\\
&&=-\frac{1}{30T^{2}\text{Vol} M}\int\sqrt{g}dx^{2}\bigg(\begin{matrix}-\frac{1}{4}R^{2} g^{ij}+\nabla^{i}\nabla^{j}R\\-g^{ij}\Delta R\end{matrix}\bigg)\partial_{t}g_{ij},\notag\\
\end{eqnarray}
 where $\Delta\equiv g^{ij}\nabla_i\nabla_j$ and we have used the fact that in two dimensions $R_{ij}=\frac{R}{2}g_{ij}$. Now the condition on positivity of $\frac{\partial S}{\partial t}$ yields the choice of the flow (without the volume constraint) :
\begin{eqnarray}
\partial_{t}g_{ij}=\frac{R^{2}}{4}g_{ij}-\nabla_{i}\nabla_{j}R+g_{ij}\Delta R.
\end{eqnarray}
To impose the volume constraint along with the requirement of the positivity of the change in entropy, we can follow \cite{Hamilton} to define the \emph{normalised flow} as 
\begin{eqnarray}
\partial_{t}g_{ij}=\frac{R^{2}-\langle R^{2}\rangle}{4}g_{ij}-\nabla_{i}\nabla_{j}R,
\end{eqnarray}
where the mean  of $R^2$ is defined generally as,
\begin{equation}
	\langle R^{2}\rangle={\int\sqrt{g}d^dx R^2\over \int\sqrt{g}d^dx}.
\end{equation}
With this flow, one can also check that 
\begin{eqnarray}
\int\sqrt{g} d^{2}x g^{ij}\partial_{t}g_{ij}=0,
\end{eqnarray}
while 
\begin{eqnarray}
\frac{\partial S_{Gibbs}}{\partial t}\approx&&\frac{1}{30T^{2}\text{Vol}M}\int \sqrt{g}dx^{2}\bigg[\frac{R^{2}}{8}(R^{2}-\langle R^{2}\rangle)\notag\\
&&-\Delta R{\langle R^2\rangle\over 4}+(\nabla^{i}\nabla^{j}R)(\nabla_{i}\nabla_{j}R)-(\Delta R)^{2}\bigg]\notag\\
&&=\frac{1}{240T^{2}}\left(\langle R^{4}\rangle-\langle R^{2}\rangle^{2}\right)>0,
\end{eqnarray}
where we get rid of the last three terms in the first line using Stokes' theorem. The change of Ricci Scalar can also be computed as
\begin{eqnarray}
\partial_{t}R=-R^{ij}\partial_{t}g_{ij}+\nabla^{i}\nabla^{j}\partial_{t}g_{ij}-\Delta g^{ij}\partial_{t}g_{ij},
\end{eqnarray}
which gives 
\begin{eqnarray}
\partial_{t}R=-\frac{R}{4}(R^{2}-\langle R^{2}\rangle)-\frac{1}{2}(\nabla_{i}R)(\nabla^{i}R).
\end{eqnarray}
We can check the behaviour of the above equation under a perturbation from the mean scalar curvature, which is fixed ($4\pi E/\text{Vol }M$), as $R=\langle R\rangle +\sigma$, generating the relation
\begin{eqnarray}
\partial_{t}\sigma=-\frac{\langle R\rangle^{2}}{2}\sigma-\frac{1}{2}\nabla_{i}\sigma\nabla^{i}\sigma-\frac{\langle R\rangle}{4}(3\sigma^{2}-\langle \sigma^{2}\rangle)+O(\sigma^{3}).\notag\\
\end{eqnarray}
If we consider only the linear order term, then the perturbation exponentially falls off. Hence the metric goes to that of $M$, if the perturbation is small. 

In the generic case, we have the relation due to \cite{Singer} as $Q=\frac{5}{2}R^{2}-R_{ij}R^{ij}+R^{ijkl}R_{ijkl}$. Therefore one can check that in any dimension $d$, it is possible to get a geometric flow implying $\frac{\partial S_{Gibbs}}{\partial t}>0$. Although as one increases the dimension the geometric flow becomes more and more complicated. For example, In $d=3$, $R^{ijkl}R_{ijkl}=4R_{ij}R^{ij}-\frac{3}{2}R^{2}$, rendering the relation $Q=3 R_{ij}R^{ij}+{3\over 2} R^2$. Following the same steps as before, in this case 
\begin{eqnarray}\label{eq:higherderi}
	\mathcal{N}&&{\partial S_{Gibbs}\over \partial t}={\partial\over \partial t}\int \sqrt{g}dx^3\left(\frac{5}{2}\langle R\rangle R-3 R_{ij}R^{ij}-{3\over 2} R^2\right)\notag\\
\end{eqnarray}
where $\mathcal{N}=180 T^2 \text{Vol}M$. This equation is exactly the classical action for higher derivative gravity \cite{Stelle:1977ry}, albeit for three dimensional situation with specified values of the coupling constants. After taking the variation with $t$, one can write
\begin{eqnarray}
	&&4\mathcal{N}{\partial S_{Gibbs}\over \partial t}=\notag\\
	&&\int\sqrt{g}dx^3\left(\begin{matrix}10\langle R\rangle \left(Rg^{ij}-2R^{ij}\right)-6 R_{ab}R^{ab}g^{ij}\\
	-24 g^{jb}g^{ik}R_{abke}R^{ea}+12\Delta R^{ij}\\ +18g^{ij}\Delta R+12R R^{ij}-3R^2g^{ij}\\-24 \nabla^i\nabla^j R\end{matrix}\right)\partial_{t}g_{ij}.\notag\\
\end{eqnarray}
Therefore we have the flow equation as (without the volume constraint):
\begin{eqnarray}
	\partial_t g_{ij}&=&10\langle R\rangle\left(2 R_{ij}-Rg_{ij}\right)+6 g_{ij}R_{ab}R^{ab}+24 \nabla_i\nabla_j R\notag\\
	&&-12 R R_{ij}+24 g_{ik}g_{jb}R^{abke}R_{ea}-12\Delta R_{ij}\notag\\
	&&+3 R^2g_{ij}-18 g_{ij}\Delta R.
\end{eqnarray}
Although the flow gets complicated as the number of dimensions increase, one can consider the flow for maximally symmetric spacetimes in any dimension, where the calculations are simple enough to track. For a maximally symmetric $d$-dimensional space, we have
\begin{equation}
	R_{ij}={R\over d}g_{ij};\quad R_{ijkl}={R\over d(d-1)}\left(g_{jk}g_{il}-g_{lj}g_{ik}\right).
\end{equation}
Hence, $Q$ turns out to be 
\begin{equation}
	Q=\mathcal{C}R^2;\quad \mathcal{C}={5\over 2}-{1\over d}+{2\over d(d-1)}.
\end{equation}
Following the same footsteps as above one can then write the flow equation as
\begin{eqnarray}
	\partial_tg_{ij}&=&5\langle R\rangle R\left({1\over d}-{1\over 2}\right)g_{ij}+\mathcal{C}R^2\left({2\over d}-{1\over 2}\right)g_{ij}\notag\\
	&&-2\mathcal{C}\nabla_i\nabla_j R+2 \mathcal{C}\Delta R g_{ij}
\end{eqnarray}
The volume constraint can be imposed in similar manner as in $d=2$.

\subsubsection{Open Manifold}\label{subsec:OpenMC}
Let us now consider the Gibbs entropy function for the partition function in the case of a $d$ dimensional manifold with a codimension one boundary as established in \ref{weyl3}. Take the $d$ dimensional space to be a flat Euclidean manifold for simplicity. In that case
\begin{eqnarray}
Z&&=\frac{T^{\frac{d}{2}}\text{Vol M}}{(4\pi)^{\frac{d}{2}}}-\frac{T^{\frac{d-1}{2}}\text{Vol }\partial\text{M}}{4(4\pi)^{\frac{d-1}{2}}}-\frac{T^{\frac{d}{2}-1}\int\sqrt{g'}dx^{d-1}H}{6(4\pi)^{\frac{d}{2}}},\notag
\end{eqnarray}
which makes the Gibbs entropy function to be
\begin{eqnarray}
&&S_{Gibbs}\approx\notag\\
&&\frac{\partial}{\partial T}\bigg[T\ln \bigg(\frac{T^{\frac{d}{2}}\text{Vol M}}{(4\pi)^{\frac{d}{2}}}\bigg)-\frac{T^{\frac{1}{2}}\sqrt{4\pi}\text{Vol }\partial M}{4\text{Vol M}}-\frac{\int\sqrt{g'}dx^{d-1}H}{6 \text{Vol M}}\notag\\
&&-\frac{1}{2}\bigg(\frac{\sqrt{4\pi}\text{Vol }\partial M}{4\text{Vol M}}\bigg)^{2}-\frac{T^{-\frac{1}{2}}\sqrt{4\pi}\text{Vol }\partial M\int\sqrt{g'}dx^{d-1}H}{24 (\text{Vol M})^{2}}\bigg].\notag \\
\end{eqnarray}
For simplicity, let us denote $h\text{Vol }\partial M=\int \sqrt{g'}dx^{d-1}H$. Hence 
\begin{eqnarray}\label{SGibbsdt_open}
&&\frac{\partial S_{Gibbs}}{\partial t}\approx-\frac{T^{-\frac{1}{2}}\sqrt{4\pi}}{8\text{Vol M}}\int \sqrt{g'}dx^{d-1}\frac{1}{2}g'^{ij}\partial_{t}g'_{ij}\notag\\
&&+\frac{T^{-\frac{3}{2}}\sqrt{4\pi}\text{Vol }\partial M}{48 (\text{Vol M})^{2}}\int \sqrt{g'}dx^{d-1}(h\frac{1}{2}g'^{ij}\partial_{t}g'_{ij}\\
&&\quad\quad\quad\quad\quad+\frac{1}{2}g'^{ij}\partial_{t}g'_{ij}H+\partial_{t}H).\notag
\end{eqnarray}
If we consider the leading order term only which is of the order $O(T^{-\frac{1}{2}})$, then the analysis is quite similar to the microcanonical case discussed earlier. Since the volume of the manifold $M$ is kept fixed, we can employ the volume preserving Mean curvature flow of \cite{Huisken+1987+35+48}, under which $\text{Vol }\partial M$ is a monotonically decreasing function.  A generalization of this can be constructed, following similar steps as \cite{Huisken+1987+35+48}, to satisfy the volume constraint as well: let $\vec{F}(\vec{x},t)$ be a family of d-dimensional hypersurfaces $M$ smoothly embedded in $\bR^{n+1}$, then define the evolution equation as
\begin{eqnarray}
\partial_{t}\vec{F}=N\vec{\nu}+N^{i}\frac{\partial\vec{F}}{\partial x_{i}},
\end{eqnarray}
where $\vec{\nu}$ is a normal vector to the surface, and $N$ and $N^{i}$ are functions on $M$.  The evolution of the metric $g'_{ij}$ is then given by :
\begin{eqnarray}
&&\partial_{t}g'_{ij}=\partial_{t}\bigg(\frac{\partial\vec{F}}{\partial x_{i}}.\frac{\partial\vec{F}}{\partial x_{j}}\bigg)\\
%&&=\bigg(\frac{\partial}{\partial x_{i}}(N\vec{\nu}+N^{l}\frac{\partial\vec{F}}{\partial x_{l}}).\frac{\partial\vec{F}}{\partial x_{j}}\bigg)+\bigg(.\frac{\partial\vec{F}}{\partial x_{i}}.\frac{\partial}{\partial x_{j}}(N\vec{\nu}+N^{l}\frac{\partial\vec{F}}{\partial x_{l}})\bigg)\notag\\
=&&Nh_{ij}+\frac{\partial}{\partial x_{i}}N^{l}g'_{lj}+N^{l}\Gamma_{il}^{k}g'_{kj}+Nh_{ij}+\frac{\partial}{\partial x_{j}}g'_{il}+N^{l}\Gamma_{jl}^{k}g'_{ik}.\notag
%&&=2N h_{ij}+g'_{lj}\nabla_{i}N^{l}+g'_{li}\nabla_{j}N^{l})\notag\\
%&&=2N h_{ij}+(\nabla_{i}N_{j}+\nabla_{j}N_{i}).
\end{eqnarray}
Which can be simply written as 
\begin{equation}\label{ADM1}
	\partial_{t}g'_{ij}=2N h_{ij}+(\nabla_{i}N_{j}+\nabla_{j}N_{i}),
\end{equation}
with the volume constraint 
\begin{eqnarray}
\frac{d\text{Vol} M}{dt}=\int \sqrt{g'}dx^{d-1}N=0.
\end{eqnarray}
The choice $N=h-H$ and $N^{l}=0$ reduces this to the volume preserving Mean curvature flow. A careful look at \eqref{ADM1} reveals that one can consider $g'_{ij}$ as the \emph{dynamic variable} that evolves with the parameter ``$t$", with $N$ and $N^i$ denoting the difference between the hypersurfaces defined through $g'_{ij}$ and the displacement of points in the hypersurface. One should therefore note the similarity of the above equation \ref{ADM1} to the ADM formalism of gravity\cite{Arnowitt:1962hi}. To further explore the relevance, one can work out the change in the second fundamental form $h_{ij}$ as
\begin{eqnarray}\label{ADM2}
\partial_{t}h_{ij}&=&-\nabla_{i}\nabla_{j}N+Nh_{jm}h_{i}^{m}\\
&&+h_{jl}\frac{\partial}{\partial x_{i}}N^{l}+h_{il}\frac{\partial}{\partial x_{j}}N^{l}+N^{l}\frac{\partial}{\partial x_{l}}h_{ij}.\notag
\end{eqnarray}
Hence the identification of $N$ to the lapse and $N_{i}$ to the shift function and the parameter $t$ to time is immediate. Now keeping till $O(T^{-\frac{1}{2}})$ in \ref{SGibbsdt_open}, the change of entropy with time becomes
\begin{eqnarray}
\frac{\partial S_{Gibbs}}{\partial t}\approx -\frac{T^{-\frac{1}{2}}\sqrt{4\pi}}{8 \text{Vol} M}\int\sqrt{g'}dx^{d-1}NH.
\end{eqnarray}
The choice $N=(h-H)$, preserves the enclosed volume, and gives the condition that the volume of the boundary manifold continuously decreases with time, as the entropy increases with time.

If on the other hand, we had not taken the $d$ dimensional space to be  a flat Euclidean manifold, then the free energy $\ln Z$ will be 
\begin{eqnarray}
&&\ln \bigg(\frac{T^{\frac{d}{2}}\text{Vol M}}{(4\pi)^{\frac{d}{2}}}\bigg)-\frac{T^{-\frac{1}{2}}\sqrt{4\pi}\text{Vol }\partial M}{4\text{Vol M}}\notag\\
&&+\frac{\int \sqrt{g}dx^{d} R-\int\sqrt{g'}dx^{d-1}H}{6T \text{Vol M}}+O(T^{-\frac{3}{2}}).
\end{eqnarray}
One can recognize the third term as the Euclidean Einstein-Hilbert action with the Gibbons-Hawking boundary term. It is interesting to note, imposition of the constraints $\text{Vol} M=\text{constant}$ and $\text{Vol}\partial M=\text{constant}$, and extremization of the free energy provides a connection similar to \cite{PhysRevLett.130.221501}. We intend to look at these directions in the future. 
\section{Discussion}\label{sec:discussion}
In this paper we started by considering the eigenvalues of the Laplace-Beltrami operator as the energies of some system, using which we define microcanonical and canonical statistical ensembles. In both cases we studied the resulting geometric flows defined through the condition ${dS\over dt}>0$ or the second law of thermodynamics. For the microcanonical ensemble, we found that the area preserving curve shortening flows will naturally satisfy the second law for a two dimensional plane with a compact boundary (one dimensional). A simple higher dimensional generalisation leads to volume preserving Yamabe-type flow and the Mean curvature flow. It is also shown that by choosing a flow closely related to gradient flow introduced by Perelman, there is equivalence between the positivity of the derivative of Perelman $\mathcal{F}$ function and the monotonic increment of entropy. 

The positivity of entropy function in the canonical case is more interesting because of its natural relation with the curvature dependent terms which are absent in the asymptotic formula of Weyl. We separately studied the geometric flows arising in this case for open and closed manifolds. Geometric flows arising in the closed manifold seem to become more and more complicated as one increases the dimension of the manifold. We showed how one can derive the flow for general dimensions but have not studied the short-time stabilities of these flows, which we will study in a separate manuscript. 

Interestingly, considering open manifolds for the canonical ensemble we observed a close connection with the ADM formalism rendering the parameter $t$ that labels the flow direction to be the time direction itself. In future we plan to explore this avenue in more detail.

\begin{acknowledgments}
We are thankful to Prof. Debashis Ghoshal and Prof. Koushik Ray for valuable discussions. PD would like to thank Arpan Saha and Arpita Mahata, with whom the project was initially conceived. The work of AC is supported by the European Union’s Horizon 2020 research and innovation programme under the Marie Sk\l{}odowska Curie grant agreement number 101034383.
\end{acknowledgments}

\bibliographystyle{apsrev4-2}
\bibliography{apssamp.bib}% Produces the bibliography via BibTeX.

\end{document}